Content-aware rankings: a new approach to rankings in scholarship

Sean C. Rife[1,2]

Joshua M. Nicholson[1]

Beatriz Bosques [1]

Domenic Rosati[1,3]

Ashish Uppala[1]

Igor A. Osipov [1]

Corresponding author: Sean C. Rife (srife@researchsolutions.com)

[1] Research Solutions, Inc.
[2] Murray State University, Murray, KY USA
[3] Dalhousie University, Halifax, Nova Scotia, CA

**Abstract**

Entity rankings (e.g., institutions, journals) are a core component of academia and related industries. Existing approaches to institutional rankings have relied on a variety of data sources, and approaches to computing outcomes, but remain controversial. One limitation of existing approaches is reliance on scholarly output (e.g., number of publications associated with a given institution during a time period). We propose a new approach to rankings – one that relies not on scholarly output, but rather on the type of citations received (an implementation of the Scite Index). We describe how the necessary data can be gathered, as well as how relevant metrics are computed. To demonstrate the utility of our approach, we present rankings of fields, journals, and institutions, and discuss the various was Scite's data can be deployed in the context of rankings. Implications, limitations, and future directions are discussed.

**Introduction**

Institutional rankings have become significant benchmarks of success for universities globally. These rankings are not merely indicators of quality but have evolved to influence institutional behaviors and practices and serve as a technique to project a curated image externally while also reflecting internal operations within organizations (Trinidad et al., 2023; Waheeduzzaman, 2007). Rankings have the power to shape what behaviors and practices are considered legitimate within institutions (Kehm, 2013). They have set standards for higher education policymaking and institutional strategies, impacting funding, attracting students, and shaping faculty recruitment (Erkkilä & Piironen, 2020). A wide variety of factors – including data from surveys, sustainability indices, test scores, grant funding, and award receipts – are used in institutional rankings. One common indicator of success is research activity, which can be operationalized in many ways, including the number of papers and presentations linked to an institution through author affiliations. Such research output can itself be assessed by measuring the number of citations to works associated with an institution.

Despite their widespread use, institutional rankings – most prominently the Times Higher Education, QS World University Rankings, and the Centre for Science and Technology Studies Leiden Rankings – have faced significant criticism due to their methodological shortcomings, lack of transparency, and the limited scope of their metrics (Fauzi et al., 2020; Van Raan, 2005; Vernon et al., 2018)[4]. One of the major criticisms is the over reliance on a small number of performance indicators, such as research outputs and reputational surveys, along with the weighting assigned to these metrics, which tends to favor well-established, research-intensive institutions (Hazelkorn, 2007; Selten et al., 2020). This creates a cycle where top ranked universities maintain their high positions due to factors like peer reputation and citation counts, regardless of the true quality or societal impact of their research (Bellantuono et al., 2022; Kayyali, 2023). Furthermore, rankings often fail to capture key dimensions of academic performance, such as teaching quality, community engagement, or regional and institutional diversity (Fauzi et al., 2020; Marginson, 2007). This has led to calls for more comprehensive and transparent evaluation methods that take into account a wider range of academic activities and the broader societal impact of research (Adler & Harzing, 2009).

---

[4] These concerns apply to all reports discussed in the present paper, but arguably less so with respect to newer, open versions of the CSTS Leiden Rankings, which are more transparent and easily audited due to their use of data from OpenAlex (see Waltman, 2024) rather than the proprietary data provided by Web of Science.

The traditional reliance on bibliometric indicators such as citation counts to measure research activity also faces significant limitations. While counting citations associated with institutions or journals is a simple, intuitive way of measuring scholarly impact, this approach is limited. As we have argued elsewhere, (e.g., Nicholson et al., 2021), simple counts of citations obscure their most important aspect: content and intent. The tendency to treat all citations as equal disregards the fact that citations may represent different forms of scholarly engagement since some may signify support for a theory, while others could show disagreement or even critique (Lamers et al., 2021). Moreover, citation metrics are susceptible to manipulation through practices like self-citation which inflate an institution's ranking without necessarily reflecting meaningful contributions to research (Buschman & Michalek, 2013). Surely, if publication records and citations are to meaningfully contribute to any institutional rankings scheme, the type of citations an institution receives is relevant.

Here, we present a new approach to using citations in institutional rankings: one that measures scholarly output not just by the number of citations a given entity has received, but the *content* of the citations themselves. In addition to describing our methodology, we also provide an example of how it may be applied to three types of entities: institutions, journals, and fields.

**Using Scite to create a Content-Aware Metric**

The present work uses data from Scite, a large, curated collection of scholarly citations and their associated text (i.e., what one paper says about another). At the time of this writing, the Scite database contains over 1.4 billion citation statements – defined as the text surrounding a citation – from over 38 million separate publications. Scite ingests papers from a variety of sources, including open access publications and through indexing agreements with publishers. It then identifies the in-text citations and extracts the text surrounding it, and links it to the published work being referenced. Finally, the text of the citation – known as the citation statement – is classified into one of three types: supporting (e.g., "our results are consistent with," "we successfully replicate"), mentioning (e.g., statements related to theory, method, or implications; statements that do not directly address an empirical claim), or contrasting (e.g., "our findings do not support," "we failed to replicate"). Detailed information about how Scite ingests and classifies citation statements can be found in Nicholson et al. (2021).

To aggregate and communicate the type of citations an entity has received, we developed the Unweighted Scite Index (USI), defined simply as the number of supporting citations the entity has received divided by the sum of the supporting and contrasting citations it has received:

$$USI = \frac{supporting}{supporting + contrasting}$$

This produces a ratio of supporting citation statements to all valanced citation statements - a value ranging 0-1, with larger values representing more support compared to fewer contrasting statements. This approach builds on previous work by Nicholson and Lazebnik (2015), who proposed a similar metric that quantifies the veracity of a scientific claim. Across all of Scite's data at the time of this writing, journals have lifetime SI values ranging from .63 to 1, while institutions have SI values ranging from .60 to 1.

Scite Index (SI) provides insight unavailable from existing metrics, as it is based on the content of citations rather than their simple existence. Interestingly, as shown in Figure 1, SI is uncorrelated with the most prominent metric of scholarly productivity, the Impact Factor (r = .03; see Figure 1).

*Figure 1: Relationship between Unweighted Scite Index (journals) and Impact Factor*

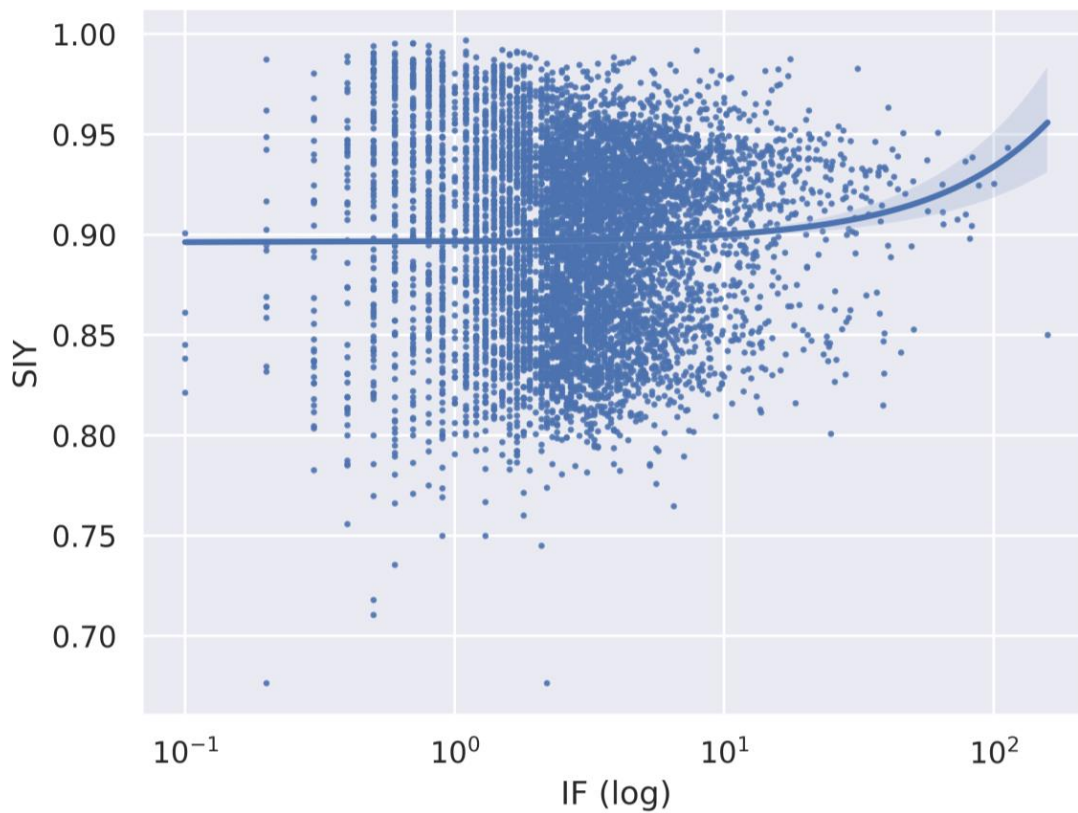

While the Unweighted Scite Index provides an accessible metric of replicability, it does not tell the full story, as only two of the three citation types Scite has identified are included (mentioning cites are not present). This results in entities receiving a small number of overall citations that happen to be supporting receiving the highest USI values (often 1 when liberal filtering is applied to lifetime USI values). To incorporate the totality of citations an entity receives into an USI value, we can compute an Impact-Weighted SI value (referred hereafter as simply SI or Scite Index), which multiplies the number of references (in the formulas denoted as citations) received by its exponentiated SI value and takes the natural logarithm of the result:

$$SI = \log((citations)(USI^2))$$

The complete formula being:

$$SI = \log\left(citations\left(\left(\frac{supporting}{supporting + contrasting}\right)^2\right)\right)$$

This applies a penalty to entities with lower USI scores – the extent of the penalty being determined by the exponent applied to the USI. This representation therefore incorporates the totality of information on an entity, but is less easily interpreted, as it has an undefined range. Therefore, we report both the weighted and unweighted SI values in order to provide data that is both easily interpretable and complete.

**Creating Rankings using USI and SI**

Having established a method of measuring scholarly impact that is based on the content of citations, we now turn to the computation of that metric and its application to entities – specifically, institutions, journals, and fields. This process involves aggregating locating citations over a given period, as well as tying those citations to the entity in question. For some entities (e.g., journals and fields), this linking task is straightforward, as the needed data is readily available in article metadata. For other entities – in particular, institutions – this task is more complicated, as individual publications are tied to a given entity through authors, who may (a) have the same name as other authors, and (b) change institutions throughout their scholarly career. As such, authors must be disambiguated in order to be properly tied to a given institution. We should also note that, as other entities (e.g., countries, regions) are tied to papers through institutions, there is a large number of entities beyond institutions themselves that rely on author disambiguation.

For rankings in the present paper, we address this issue by using a curated dataset for cited articles that has been vetted for accuracy: the Leiden 2023 Open Rankings data. This dataset uses the OpenAlex platform to link publications to institutions and has approximately 93% accuracy in author affiliations.

For all three entity types, we examine citations in works published in the year 2024. This limitation is on the *citing* works only – the publications being cited may have been published in any year (Scite's data contain publications from as early as the fifteenth century). As interest in various topics waxes and wanes, even older publications that are seminal works may receive a heightened level of interest in recent years.

When institutional rankings are sorted by the weighted SI value (see Table 1), prominent names are surfaced, reflecting the high amount of publication output associated with research-intensive institutions. A similar pattern can be observed when sorting journals by their weighted SI value (see Table 2), with prominent publications at the top of the list, due to these journals receiving a large number of references.

*Table 1: Institutional data sorted by SI scores (top 10 institutions)*

| Institution | Supporting | Mentioning | Contrasting | USI | SI |
|---|---|---|---|---|---|
| Harvard University | 34,516 | 989,505 | 3,776 | 0.9 | 6.24 |
| Stanford University | 15,572 | 469,817 | 1,752 | 0.9 | 5.94 |
| Massachusetts Institute of Technology | 12,928 | 408,028 | 992 | 0.93 | 5.90 |
| University College London | 17,248 | 444,902 | 2,117 | 0.89 | 5.89 |
| University of Toronto | 14,642 | 422,925 | 1,871 | 0.89 | 5.88 |
| University of Oxford | 15,432 | 411,997 | 1,743 | 0.9 | 5.87 |
| University of the Chinese Academy of Sciences | 10,676 | 303,849 | 937 | 0.92 | 5.87 |
| University of Michigan | 12,482 | 355,124 | 1,411 | 0.9 | 5.83 |
| University of Cambridge | 13,743 | 366,057 | 1,391 | 0.91 | 5.83 |
| Johns Hopkins University | 13,874 | 395,169 | 1,881 | 0.88 | 5.82 |

*Table 2: Publication data sorted by SI scores (top 10 journals)*

| Journal | Supporting | Mentioning | Contrasting | USI | SI |
|---|---|---|---|---|---|
| Nature | 33,505 | 1,022,155 | 2,652 | 0.93 | 6.31 |
| Science | 24,214 | 808,418 | 2,054 | 0.92 | 6.21 |
| PNAS | 37,913 | 868,129 | 3,232 | 0.92 | 6.19 |
| Angewandte Chemie | 11,837 | 529,278 | 706 | 0.94 | 6.03 |
| Nature Communications | 19,874 | 487,258 | 1,547 | 0.93 | 5.95 |
| Scientific Reports | 20,415 | 387,667 | 2,186 | 0.90 | 5.88 |
| ACS Applied Materials & Interfaces | 7,666 | 324,768 | 492 | 0.94 | 5.87 |
| New England Journal of Medicine | 10,068 | 404,919 | 1,504 | 0.87 | 5.86 |
| PLOS ONE | 26,117 | 440,505 | 4,104 | 0.86 | 5.86 |
| Physical Review B | 10,599 | 199,995 | 594 | 0.95 | 5.80 |

However, when sorting by unweighted SI scores (see tables 3 and 4), an interesting pattern emerges. The highest unweighted SI scores are held by journals that publish in the harder physical sciences, applied topics, and engineering. Similarly, the institutions with the highest unweighted SI scores tend to be technical institutions. These differences by field can also be observed by plotting institutional SI scores against the fields of their associated publications (see

Figure 2), which reveals that mathematics and physics have comparatively high SI scores, while the social and life sciences have comparatively low SI scores.

*Table 3: Publication data sorted by USI scores (top 10 journals)*

| Journal | Supporting | Mentioning | Contrasting | USI | SI |
| --- | --- | --- | --- | --- | --- |
| Electronic Journal of Plant Breeding | 193 | 298 | 1 | 1.00 | 2.83 |
| Indian Journal of Weed Science | 292 | 764 | 2 | 0.99 | 3.05 |
| The European Physical Journal E | 237 | 3,651 | 2 | 0.99 | 3.93 |
| Advanced Materials Technologies | 113 | 9,779 | 1 | 0.99 | 4.37 |
| Reactive and Functional Polymers | 108 | 3,434 | 1 | 0.99 | 4.07 |
| Computer Physics Communications | 102 | 12,045 | 1 | 0.99 | 4.59 |
| Geoscience Frontiers | 101 | 2,566 | 1 | 0.99 | 4.10 |
| Journal of the Meteorological Society of Japan Series II | 97 | 1,711 | 1 | 0.99 | 3.66 |
| The Neuroscientist | 194 | 5,904 | 2 | 0.99 | 4.14 |
| Beilstein Journal of Nanotechnology | 95 | 2,532 | 1 | 0.99 | 3.80 |

*Table 4: Institutional data sorted by USI scores (top 10 institutions)*

| Institution | Supporting | Mentioning | Contrasting | USI | SI |
|---|---|---|---|---|---|
| Rzeszów University of Technology | 75 | 2,101 | 1 | 0.99 | 3.79 |
| Indian Institute of Technology Patna | 105 | 2,867 | 1 | 0.99 | 4.02 |
| Xiangtan University | 304 | 9,647 | 5 | 0.98 | 4.49 |
| Military University of Technology, Warsaw | 79 | 1,987 | 2 | 0.98 | 3.80 |
| Changchun University of Technology | 54 | 2,898 | 1 | 0.98 | 3.98 |
| Shenyang University of Technology | 63 | 1,960 | 1 | 0.98 | 3.95 |
| Moscow Institute of Physics and Technology | 363 | 10,306 | 12 | 0.97 | 4.43 |
| Academy of Scientific and Innovative Research | 598 | 21,055 | 21 | 0.97 | 4.78 |
| Xinjiang University | 226 | 6,796 | 7 | 0.97 | 4.36 |
| Shanmugha Arts, Science, Technology & Research Academy | 116 | 4,524 | 3 | 0.97 | 4.20 |

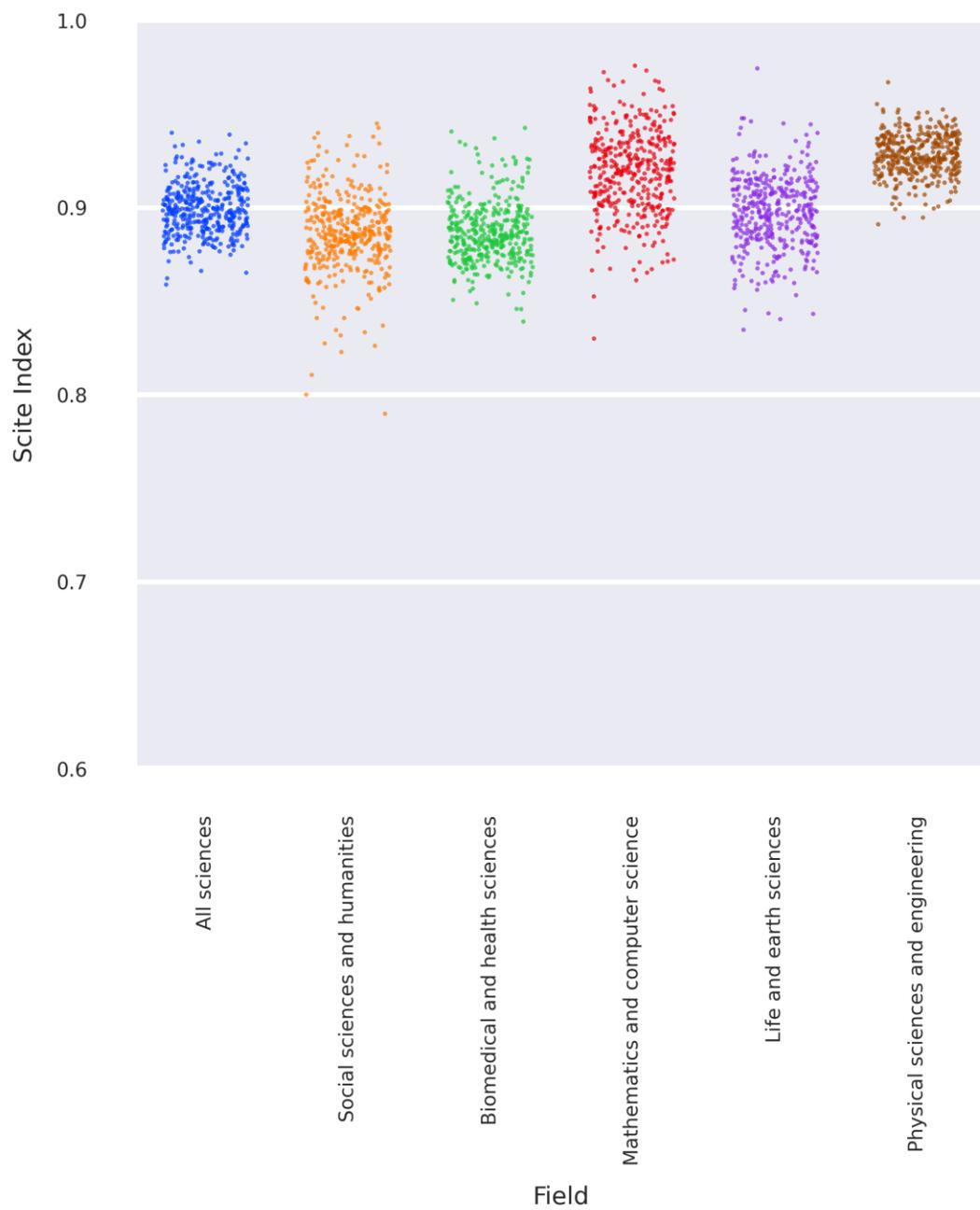

*Figure 2: Institutional Scite Index scores by field*

**Conclusions, Limitations, and Future Directions**

Using citations for institutional and journal rankings is a unique approach that results in meaningful differences when compared to other rankings reports. For example, CWTS Leiden Rankings measure scientific impact by examining the publication output of a curated list of universities. This metric represents productivity well but may be less effective in assessing scientific impact. In contrast, our approach assesses the impact a given institution has had based on how other researchers cite published research – a variable tied more closely to the dynamic interplay between scholarly publications and the gradual nature of scientific progress. Interestingly, the list of top institutions presented in the present paper varies considerably from other rankings. For example, while many of the institutions in the top 10 presented by QS Rankings (which does include citations received as part of their rankings model) are also present in Table 1 above, their positions vary considerably. These differences are due in part to the inclusion of citation context in our rankings metric rather than relying solely on a simple count of citations.

**Limitations**

Our approach to rankings has a number of limitations, the most significant of which is our reliance on the full text of articles in the computation of Scite Index scores. While Scite ingests all available open access content, author versions, as well as closed-access papers through indexing agreements with publishers, this approach does exclude closed-access papers from publishers with whom Scite does not have an indexing agreement. However, the weighted Scite Index score is based also on the number of references a paper has received – data that are often available as metadata, and thus obtainable regardless of whether an indexing agreement is in place. As the number of publishers with whom Scite has indexing agreements increases, this problem will abate.

Additionally, none of the ranking models are immune to the language issue. English language publications form the foundation of all the datasets used in the key university ranking models worldwide. While it reflects the fact that scholarship is utilizing English as a primary mean of communication, there is notable research data which is missing in the rankings model because of that, skewing results toward the institutions producing English-language research. Scite dataset of scholarly papers is similarly shaped by this global trend, but is addressing this problem step by step (e.g., Scite, 2022), gradually including sources in other scholarly languages.

As noted earlier, while journal rankings are straightforward (given that our ability to link citations to specific works is quite good), institutional rankings are more complicated, as they rely on correctly identifying authors and tying them to institutions. Since author names can change or be shared by multiple individuals, as well as change from one institution to another, this is a challenge. Advancements in author disambiguation will likely improve the accuracy and coverage of author-institution links in the future.

**Future Directions**

Given that our approach to rankings is relatively new, there is much work to be done in terms of its validation and expansion. Further comparisons with established ranking schemes (especially going forward, as new data are generated with each iteration of ranking reports) will be especially informative and may warrant changes to how the Scite Index is computed. For example, changing how the unweighted SI is exponentiated is a straightforward way of adjusting how much importance is given to supporting and contrasting citations compared to references alone. Additionally, it may be useful to incorporate the third type of citation Scite generates – a mentioning citation – into the formula for SI, either replacing or augmenting the raw reference count that is currently part of the SI equation.

There are also additional, novel approaches to entity rankings that can be constructed by combining SI data with existing formulas. For example, the h-index can be easily modified to be based solely on supporting citations. An "integrity index" could be computed to highlight authors who routinely cite their previous work in a contrasting manner, indicating that they are open to revising their position on key theories and/or findings. In short, we believe there is a bright future for a system of rankings – as well as other scholarly metrics – that incorporates the content of citations rather than there mere existence.